\documentstyle[12pt,epsf,fleqn]{elsart}

\def \ni{\noindent}

\def \be {\begin{equation}}
\def \ee {\end{equation}}

\begin{document}

\ni {\bf Corresponding author}\\
Prof. Dr. H.V. Klapdor-Kleingrothaus\\
Max-Planck-Institut f\"ur Kernphysik\\
Saupfercheckweg 1\\
D-69117 HEIDELBERG\\
GERMANY\\
Phone Office: +49-(0)6221-516-262\\
Fax: +49-(0)6221-516-540\\
email: $klapdor@gustav.mpi-hd.mpg.de$\\

\begin{frontmatter}
\title{Searching for the Annual Modulation of Dark Matter 
signal with the GENIUS-TF experiment.}

\author{C. Tomei$^1$%\footnote{On leave from Universit\`a degli Studi de
    %L'Aquila, Italy}
, A. Dietz, I. Krivosheina, }
 \protect\newline {H.V. Klapdor-Kleingrothaus}
\address{Max-Planck-Institut f\"ur Kernphysik, PO 10 39 80, 
  D-69029 Heidelberg, Germany}
\address{$^1$On leave from Universit\`a degli Studi de L'Aquila, Italy}
\date{}
\maketitle

\begin{abstract}
%\begin{center}
%\section*{Abstract}
%\end{center} 
	The annual modulation of the recoil spectrum observed 
	in an underground detector is well known as the main 
	signature of a possible WIMP signal.

	The GENIUS-TF experiment, under construction in the Gran Sasso
	National Laboratory, can search for the annual modulation of the Dark
	Matter signal using 40\,kg of naked-Ge detectors in liquid nitrogen. 
	Starting from a set of data simulated under the hypothesis 
	of modulation
	and using different methods, we show the potential of GENIUS-TF for 
	extracting the modulated signal and the expected WIMP mass and 
	WIMP cross section.
\end{abstract}
\end{frontmatter}

%\vspace{1.5cm}

\section{Introduction}
	It is generally assumed that our galaxy is embedded in a halo of dark
	matter particles (WIMPs) with energy density $\rho \simeq 0.3$ 
	GeV/cm$^3$ and velocities di\-stributed according to a
	Maxwellian distribution with parameter $v_0$ (defined as
	$\sqrt{(\frac{2}{3})}\; v_{\rm{rms}}$) and cut-off 
	velocity equal to the escape velocity 
	in the Galaxy ($v_{\rm{esc}} \simeq 650$ km/s).

	The recoil spectrum produced by WIMP-nucleus scattering in
	a target detector is expected to show the so-called  
	annual modulation effect, due to the Earth's motion around the 
	sun 
\cite{Freese}. 
	Along the year, the Earth's velocity with respect to the galactic
	reference frame ($v_E$) varies according to a cosinus law. 
	It is customary to express the Earth's velocity in units 
	of the parameter $v_0$ defined above. 
	The adimensional quantity $\eta = v_E/v_0$ shows the
	following time-dependence: 
\begin{equation} \label{eta}
 \eta(t) = \eta_0 + \Delta\eta\cos\omega(t-t_0),
\end{equation}
	where the amplitude of the modulated
	component ($\Delta\eta \simeq 0.07) $ is small compared 
	to the annual average 
	$\eta_0 \simeq 1.05 $. The period and phase of the cosinus function
	are known to be $\omega = 2\pi$/T  (T= 1 year) and 
	$t_0 \simeq 2^{nd}$ June.

	In this framework, the expected countrate of WIMP interactions can be
	written (first order Taylor approximation): 
\begin{equation} \label{signal}
S = S_0 + S_{\rm{m}}\cos\omega(t-t_0)
\end{equation}
	where $S_0 = S_{\rm{k}}[\eta_0]$ is the time-independent part 
	and $S_{\rm{m}} =
\frac{\partial S_{\rm{k}}}{\partial\eta}\Big|_{\eta_0} \Delta\eta$ 
	is the amplitude of the modulated signal.

	Both $S_0$ and $S_{\rm{m}}$ depend on the WIMP mass $m_W$ and on the
	WIMP cross section on proton $\sigma_{\rm{p}}$, as 
	well as on many astrophysical parameters 
	($v_0$, $v_{\rm{esc}}$, $\rho$)
	and some properties of the target material. 

	A calculation of the expected count rate R from WIMP interactions in a
	germanium detector as a function of the recoil energy $E_R$ can be
	performed following the formula given in ref.\cite{cross1}:
\begin{equation} \label{formula_rate}
\frac{dR}{dE_R} =\frac{\rho}{m_W} N\sigma_{\rm{Ge}}\frac{m_{\rm{Ge}}c^2}{4m^2_{\rm{red}}(m_{\rm{Ge}},m_W)v_0}\frac{g(\eta,E_R)}{\eta}\;F^2(E_R)
\end{equation}
	where:
\begin{displaymath}
g(\eta,E_R) = \left\{ \begin{array}{ll}
{\rm{erf}}(\xi + \eta)-{\rm{erf}}(\xi - \eta) -
\frac{4}{\sqrt{\pi}}e^{-z^2} &
\textrm{if $\xi \le z-\eta$}\\
\\
{\rm{erf}}(z)-{\rm{erf}}(\xi - \eta) -
\frac{2}{\sqrt{\pi}}(z + \eta - \xi)e^{-z^2} &
\textrm{if $z-\eta \le \xi \le z+\eta$}\\
\\
0 & 
\textrm{if $\xi \ge z+\eta$}\\
\end{array} \right.
\end{displaymath}
	and the variables $\xi$, $\eta$ and $z$ are listed in Table \ref{tab1} 
	together with the values of the parameters used in the calculation.
	The WIMP cross section on germanium $\sigma_{\rm{Ge}}$ can be easily
	related to the correspondant cross section on proton; in the case of 
	spin-independent interaction (SI), the conversion formula is (see
\cite{cross1} 
	and 
\cite{Cebrian}):
\begin{equation} \label{conv}
\sigma_{\rm{Ge}} =
\frac{m^2_{\rm{red}}(m_{\rm{Ge}},m_W)}{m^2_{\rm{red}}(p,m_W)}\;c\;\sigma_p
\end{equation}
where $m^2_{\rm{red}}$ is the reduced mass and $c=A^2$. 

\begin{table}[!htp]
\begin{center}
\caption{Relevant astrophysical and detector parameters: their
  expressions and the values used in the present work.}\label{tab1}
\vspace{0.3cm}
\begin{tabular}{|c|c|c|c|}
\hline
%& & & \\
$v_{\rm{E}}$ & 232 km/s & $\xi$ & $\sqrt{\frac{M_{\rm{Ge}}E_R}{2m^2_{\rm{red}}v_0^2}}$ \\
$v_{\rm{esc}}$ & 600 km/s  &  &  \\
$v_0$ & 220 km/s & $\eta$ & $v_E/v_0$\\
$\rho$ & 0.3 GeV/cm$^{3}$ & & \\
$ M_{Ge}$ & 72.59 (uma) & $z$ & $v_{\rm{esc}}/v_0$ \\
%& & & \\  
\hline
\end{tabular}
\end{center}
\end{table}

%%%\noindent
	For the nuclear form factor $F^2(E_R)$, necessary to take into account 
	the finite size of the nucleus, we use a Besselian approximation 
	(see again ref. 
\cite{cross1}).

	Using formula (\ref{formula_rate}) and its derivative with respect to
	the quantity $\eta$, we can calculate the expected
	value of $S_0$ and $S_{\rm{m}}$ (units of
	counts/kg keV day) in our 
	GENIUS-TF experiment for different values of the WIMP mass and
	cross section, as a function of the energy released in the detector.

\begin{figure}[!htp]
\epsfysize=70mm\centerline{\epsffile{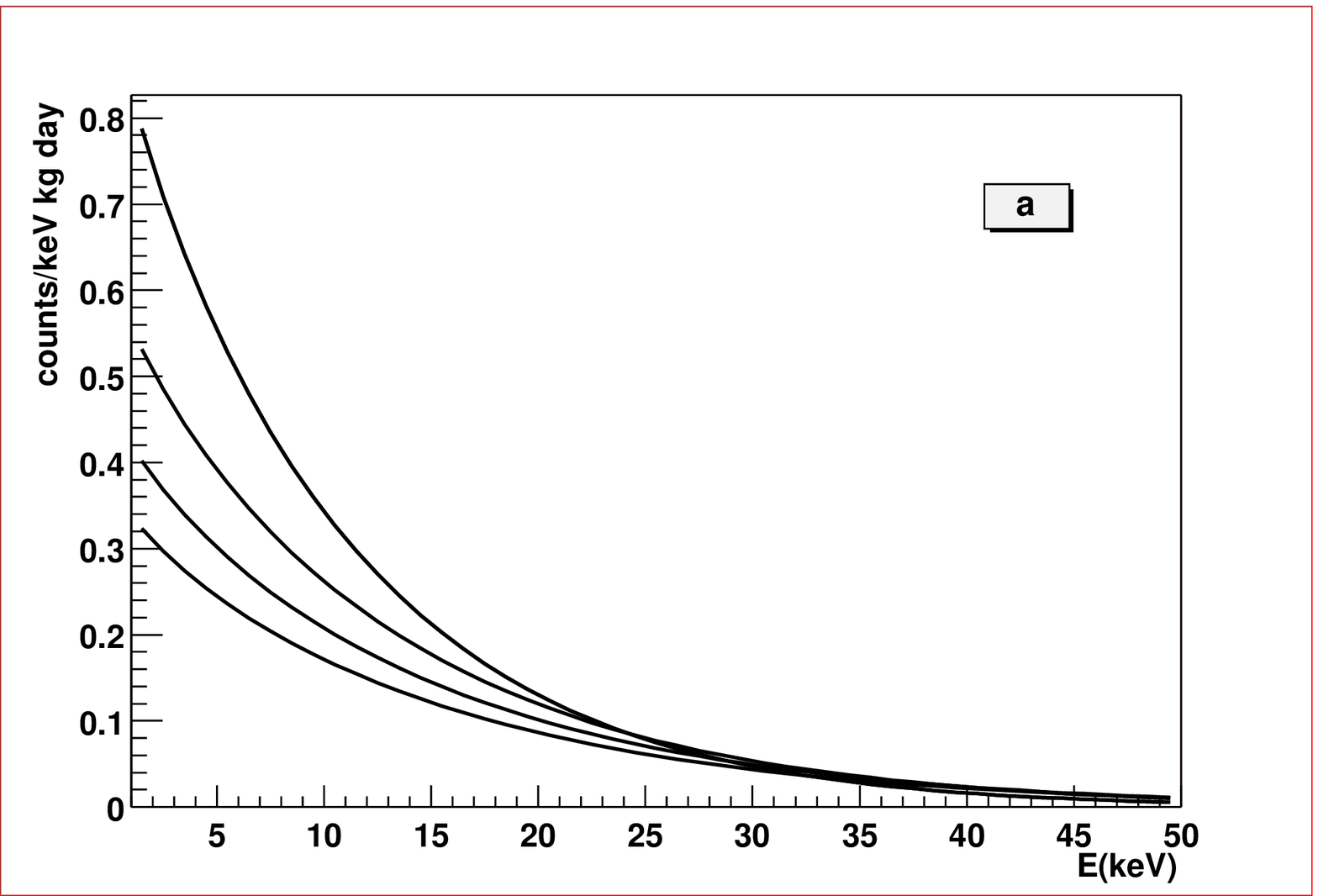}}
\epsfysize=70mm\centerline{\epsffile{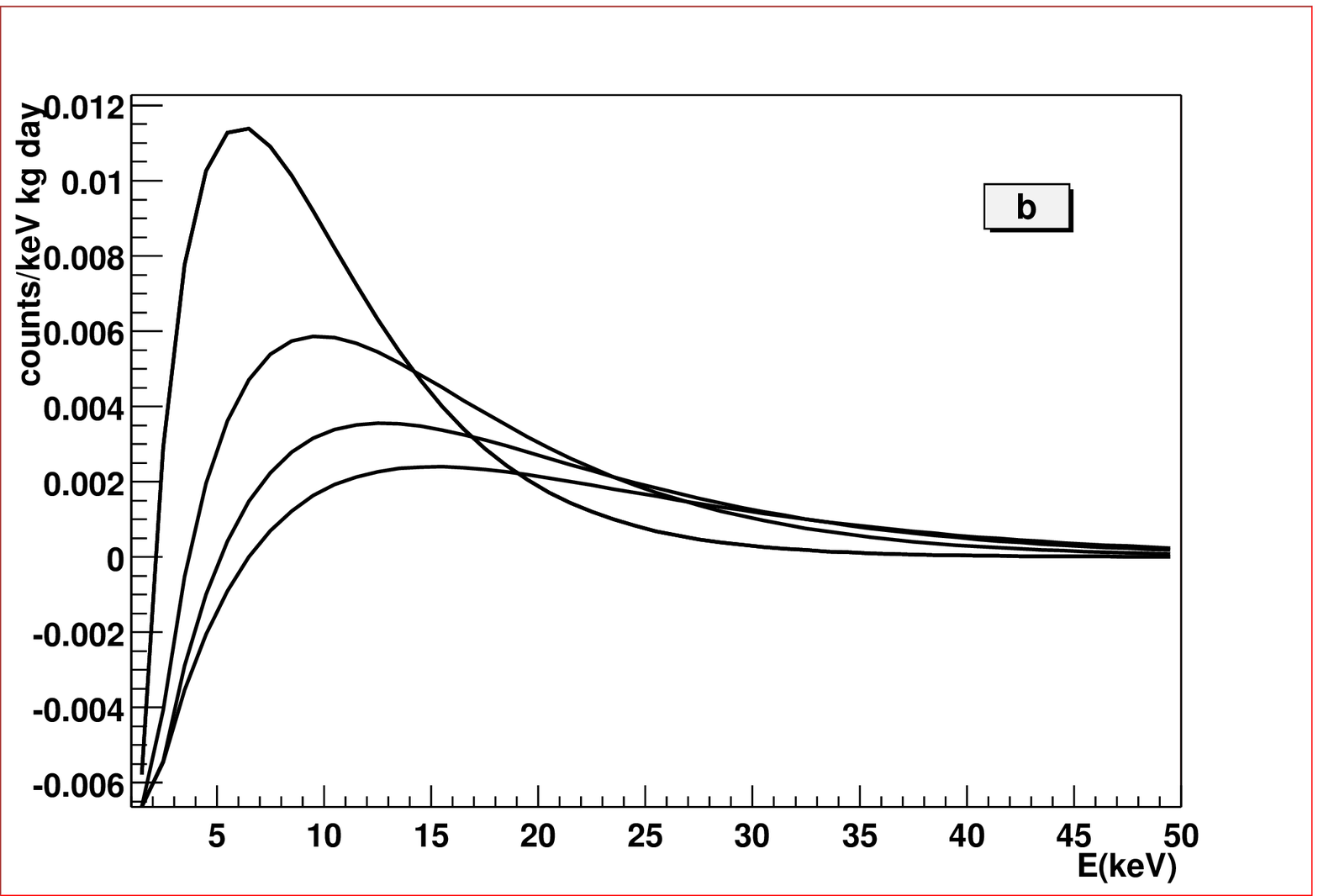}}
\caption{Expected WIMP rate in Ge for $m_W$= 40, 60, 80, 100 GeV (from top 
  to bottom) and $\sigma_{Ge} = 10^{-34} $ cm$^2$: a) time-independent 
  component of the signal ($S_0$) ; b) amplitude of the modulated
  component ($S_{\rm{m}}$).} \label{spectra}
\end{figure}

	The spectra in fig. 
\ref{spectra} have been calculated for WIMP masses
	running from 40 to 100 GeV (in steps of 20 GeV) and for a cross
	section on germanium 
	of $\sigma_{Ge} = 10^{-34} $ cm$^2$  (assuming a WIMP mass of 40 GeV 
	this corresponds to $\sigma_p = 2.63\times 10^{-5}$ pb).

	A suitable conversion law for germanium (see 
\cite{Baudis}) 
	is used to convert the recoil energy of the nucleus ($E_R$) 
	into the visible energy 
	actually released in the detector.

	As we can see from fig. 
\ref{spectra}a, 
	the time-independent component
	of the signal $S_0$ is exponentially
	decreasing with the energy masses. The amplitude of the modulated
	signal $S_{\rm{m}}$ (fig.
\ref{spectra}b) 
	is only a small fraction of the total signal (note the
	different scale of the pictures) and moreover its contribution to the
	total signal $S$ 
	can be not only positive but also negative or zero.

	Due to the small entity of the annual modulation effect, a
	sufficiently high exposure and a great stability of the experimental
	conditions over the time are required in order to detect the presence
	of a possible modulation in a set of experimental data. It is
	important to point out that two different analysis can be performed: a 
	model-independent analysis, in which only the presence of the
	modulation is looked for in the data and a model-dependent analysis,
	where, assuming a complete model framework, one allows (or excludes)  
	a region in the space of the parameters $m_W$ and $\sigma_{\rm{p}}$.
	Therefore, if a modulation signature is discovered in the experimental
	data, it is possible to extract informations 
	on the WIMP relevant quantities
	only in the framework of a given model (for example: spin-independent 
	WIMP-nucleon interactions, non-rotating halo, Maxwellian distribution
	of WIMP's velocity and so on).

%%\noindent
	So far, the DAMA experiment has reported results on 4 annual cycles of 
	observation 
\cite{DAMA} 
	and claims for a positive evidence of the annual
	modulation effect. In a model independent analysis, the probability of
	an unmodulated behaviour of the experimental rate is
	$4\cdot10^{-4}$. In the
	framework of spin-independent (SI) WIMP-matter scattering, the best fit
	values for the WIMP mass and cross section from the DAMA experiment
	are $m_W =(43^{+12}_{-9})$ GeV and 
	$\sigma_{\rm{p}} = (5.4 \pm 1.0)\cdot 10^{-6}$ pb, where 
	$\rho = 0.3$ GeV$\cdot$cm$^3$ and $v_0 = 220$ km/s have been 
	assumed.

	Since in the present work we have used these values of WIMP mass and
	cross section for our simulation, we were careful to work in
	the same framework
	of ref. 
\cite{DAMA} 
	and we have used the same values for all the astrophysical
	parameters (see Table \ref{tab1}).

%%%%%%%%%%%%%%%%%% Section 2 %%%%%%%%%%%%%%%%%%%%%%%%

\section{The GENIUS-TF experiment}

%%%%%%%%%%%%%%%%%% Section 2 %%%%%%%%%%%%%%%%%%%%%%%%

	The GENIUS-TF 
\cite{GENIUS-TF} 
	experiment is born as a test facility
	for the GENIUS 
\cite{GENIUS} 
	project. It is at present under installation at 
	the Gran Sasso National Laboratory (LNGS) and it is designed to test 
	experimentally some features for the feasibility 
	of the GENIUS experiment. 

	GENIUS-TF consists of 14 natural Ge crystals (40 kg) operated in a
	volume of 0.064 m$^3$ ultra-pure liquid nitrogen. The liquid
	nitrogen is housed by a steel vessel (0.5 mm thick) 
	inside a (0.9 m $\times$ 0.9 m $\times$ 0.9 m) 
	box of polystyrene foam, with a 5 cm thick
	inner shield of high purity Ge bricks.
	Outside the foam box there will be 10 cm of low-level copper, 30 cm of
	lead and 15 cm of borated polyethylene as shield against the natural
	radioacti\-vity of the environment.

\begin{figure}[!htp]
\epsfysize=70mm\centerline{\epsffile{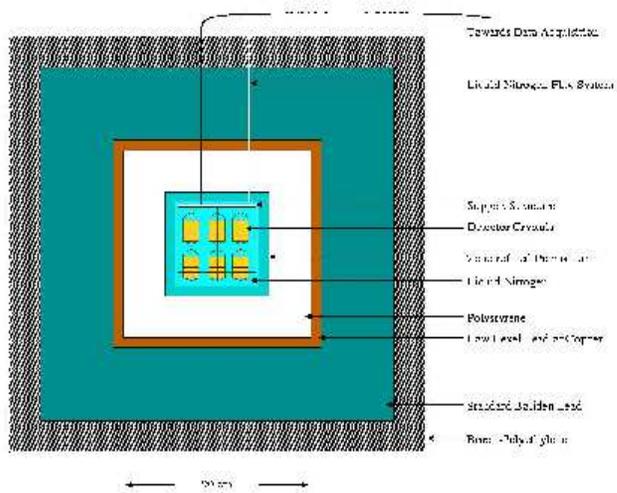}}
%%%\begin{center}
%%%\includegraphics{genius_tf.ps}
%%%\end{center}
\caption{A schematic view of the GENIUS-TF experiment.}\label{genius}
\end{figure}

%%%\noindent
	The general layout of the experiment is shown in fig.
\ref{genius}. 
	The Ge crystals are positioned in two layers on a
	holder system made of high molecular polyethylene. The signal and high 
	voltage contacts of the crystals are established trying to minimize
	the amount of material nearby the detectors, and made of stainless
	steel (about 3 g). Further details on the setup are described in the
	GENIUS-TF proposal 
\cite{GENIUS-TF}.

	Concerning the background, the aim of GENIUS-TF is to reach a level of 
\mbox{2 - 4 counts/(kg y keV)} in the energy region below 50 keV
	(corresponding to about 200 keV nuclear recoil energy). 
	This value is one order of magnitude lower than the
	actual background of the 
\mbox{Heidelberg-Moscow} experiment and two orders of
	magnitude higher than the final goal of GENIUS. 
	Careful si\-mulations of 
	the background, show that the level of counts mentioned 
	above can be reached
(\cite{Bela1}, \cite{Bela2}). 
	These simulations have been performed
	taking into account all possible sources of background (natural decay
	chains, cosmogenic activation by cosmic rays, anthropogenic
	radionuclides, contribution of neutrons and muons) and assuming
	standard values for the radioactive contaminations of different
	materials in the experimental setup. 

	In the following we assume that the background level for the
	experiment in the region below 50 keV is 
\mbox{b = 0.01 counts/(kg keV day)} 
	(corresponding to 4 counts/(kg keV y)).

	With this level of background and a detector mass of 40 kg, GENIUS 
	Test Facility can have a physics program of its own in the domain of  
	WIMP search through the annual modulation effect.\\
	It has been already shown 
\cite{Cebrian} 
	that the region of interest for the WIMP
	mass and cross section indicated by the DAMA experiment is within
	reach of many future experiments looking for the annual modulation
	signature. 

	In the specific case of a germanium detector with a
	background level of 
\mbox{0.01 counts/(kg keV day)} (the same that we assume
	for our GENIUS-TF) one can fully enter the DAMA region with exposures
	ranging from 10 to 100 kg$\cdot$year depending on the energy threshold 
	of the experiment (see fig. 1 and 2 of
\cite{Cebrian}). 
	Such exposures are achievable by the GENIUS-TF
	experiment within few years of measurement.

	In this work we follow a different approach from 
\cite{Cebrian}; 
	assumed a WIMP of a mass $m_W$ and cross section 
	$\sigma_p$ in a given theoretical framework, we simulate a set of
	experimental countrates and we show which methods can be used in the
	GENIUS-TF experiment to extract the presence of the annual modulation
	and the values of the parameters $m_W$ and $\sigma_p$.

\vspace{-0.5cm}
%%%%%%%%%%%%%%%%%% Section 3 %%%%%%%%%%%%%%%%%%%%%%%%
\section{Simulation of experimental data}
%%%%%%%%%%%%%%%%%% Section 3 %%%%%%%%%%%%%%%%%%%%%%%%

	To analize the potential of GENIUS-TF in searching for the annual
	modulation effect, we have simulated a set of experimental 
	count rates, as they would be
	recorded in our detector under the hypothesis of a WIMP
	with a given mass and cross section and in the framework of a given
	model, in the following way.
 
	Let $N_{ij}$ be the number of counts detected in the i-th 
	time bin and in the j-th energy bin;
	if there is no WIMP signal we have:
\begin{equation} \label{rate}
\langle N_{ij} \rangle = b_j M \Delta T_i \Delta E_j
\end{equation}
	where $b_j$ is the expected background in the j-th energy bin
	(as usual in units of counts/keV kg days), M is the detector 
	mass in kg, $\Delta T_i$ the
	amplitude of the time bin in days and $\Delta E_i$ the amplitude of the
	energy bin in keV.

	If, instead, we assume that WIMPs are
	contributing to the signal recorded in the detector, equation
	(\ref{rate}) becomes:
\begin{equation} \label{rate2}
\langle N_{ij} \rangle = [b_j + S_{0,j} + S_{m,j}\cos\omega(t_i-t_0)]M \Delta T_i
\Delta E_j .
\end{equation}
	As already mentioned before, we will assume $b_j$ to be constant and 
	equal to 0.01 counts/keV kg days, while for $S_{0,j}$ and $S_{m,j}$ we
	take the values calculated according to 
	(\ref{formula_rate}) for the corresponding
	energy bin. Here and in the following we assume $\Delta T_i$ = 1
	day and $\Delta E_i$ = 1 keV.

	If $N_{ij}$ is a Poisson-distributed random variable with
	mean $\mu_{ij}$ given by eq. 
(\ref{rate2}), 
	we can obtain the
	desired count rates calculating for each time and energy interval 
	the value of $\mu_{ij}$ and then randomly extracting $N_{ij}$ from
	the Poisson distribution: 
$P(N,\mu_{ij}) = e^{-\mu_{ij}}\frac{\mu_{ij}^N}{N!}$.

	Such a simulation has been performed for the energy-interval 
	(0, 50) keV, for this is the region where a possible 
	signal should be searched for 
\cite{Cebrian}. 
	For the WIMP mass $m_W$ we have chosen values around
	the best-fit result of the DAMA experiment (40, 50, 60 GeV) since we
	wanted first of all to understand
	the potential of GENIUS-TF for testing the region of parameters
	allowed, at the moment, by the DAMA experiment. 
	We have also assumed a cross section
	on proton of $5.4\cdot 10^{-6}$ pb, corresponding 
	to the best-fit value of the DAMA 
	analysis for $v_0$ = 220 km/s.

%%%%%%%%%%%%%%%%%% Section 4 %%%%%%%%%%%%%%%%%%%%%%%%
\section{Modulation analysis}

%%%%%%%%%%%%%%%%%% Section 4 %%%%%%%%%%%%%%%%%%%%%%%%

	The way to extract the modulated signal, with the proper period and
	phase, from a set of experimental data has been discussed 
	by many authors in the literature
(\cite{Freese2}, 
\cite{DAMA}, 
\cite{Cebrian}, 
\cite{DAMA2}-\cite{Sarsa}).
	In the present work we will apply 3 different methods: first we try to 
	identify, at a certain confidence level, the presence 
	of the modulation in our data, then we apply the 
	maximum likelihood method to find out the values of the WIMP's
	relevant parameters.

%%%%%%%%%%%%%%%%%% SubSection 4 %%%%%%%%%%%%%%%%%%%%%%%%

\subsection{The modulation significance}
%%%%%%%%%%%%%%%%%% subSection 4 %%%%%%%%%%%%%%%%%%%%%%%%

	Following 
\cite{Freese2} 
	and defining $S_i$ the number of counts collected 
	in the i-th day of observation integrated over a given energy interval
$$
S_i = \sum_{j=E_{i}}^{E_{f}} N_{ij}\;,
$$
	we can project out the
	modulated component of the data through the varia\-ble $r$, called
	modulation significance:
\begin{equation} \label{rest}
r=\frac{\sum_i 2 \cos\omega (t_i-t_0)S_i}
{\sqrt{2 \sum_i S_i}}
\end{equation}
	If no modulation is present in the data the variable $r$ is
	expected to be nearly Gaussian distributed with zero mean and unit
	variance. On the contrary, for experimental data modulated according
	to (\ref{signal}), the mean value of $r$ 
	will be no longer zero, but it will increase with the collected
	statistics. At the
	same time the distribution of the variable $s$:
\begin{equation} \label{sest}
s=\frac{\sum_i 2 \sin\omega (t_i-t_0)S_i}
{\sqrt{2 \sum_i S_i}} 
\end{equation}
	orthogonal to $r$, will be unaffected by the presence 
of a modulation.

	Once a value $r_0$ of $r$ has been measured in an experiment, one can
	exclude the absence of the annual modulation (with the correct period
	and phase) at a confidence level given by:
\begin{equation}
C.L.=\frac{1}{2} + \frac{1}{2}\;{\rm{erf}} \left[\frac{r_0}{\sqrt{2}}\right]
\end{equation}

\begin{figure}[!htp]
\epsfysize=70mm\centerline{\epsffile{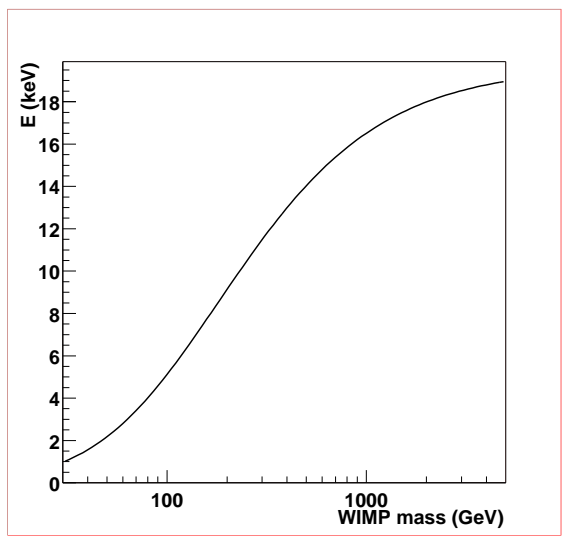}}
%%%\begin{center}
%%%\includegraphics[width=7cm,height=6cm]{cross.eps}
%%%\end{center}
\caption{Cross-over energy in Ge as a function of WIMP mass. We assume 
  $v_0$ = 220 km/s.} \label{cross}
\end{figure}

	The choice of the energy interval $(E_{i},E_{f})$ where we integrate
	the signal is important. It is well known that one should avoid
	the so-called cross-over region, i.e. the region where the
	differential spectra of June (maximum) and December (minimum)
	cross. Integrating over this region could hide a possible signal.
	The cross-over energy ($E_C$) depends on the WIMP mass and on
	the $v_0$ parameter; in case of Ge detectors $E_C$ 
	usually lies between 1 and 30 keV 
\cite{Hasenbalg}. 
	In fig. 
\ref{cross} 
	we have explicitly calculated its value as a
	function of the WIMP mass assuming as usual $v_0$ = 220 km/s. The
	result is in agreement with \cite{Hasenbalg}.

\begin{figure}[!htp] 
\epsfysize=70mm\centerline{\epsffile{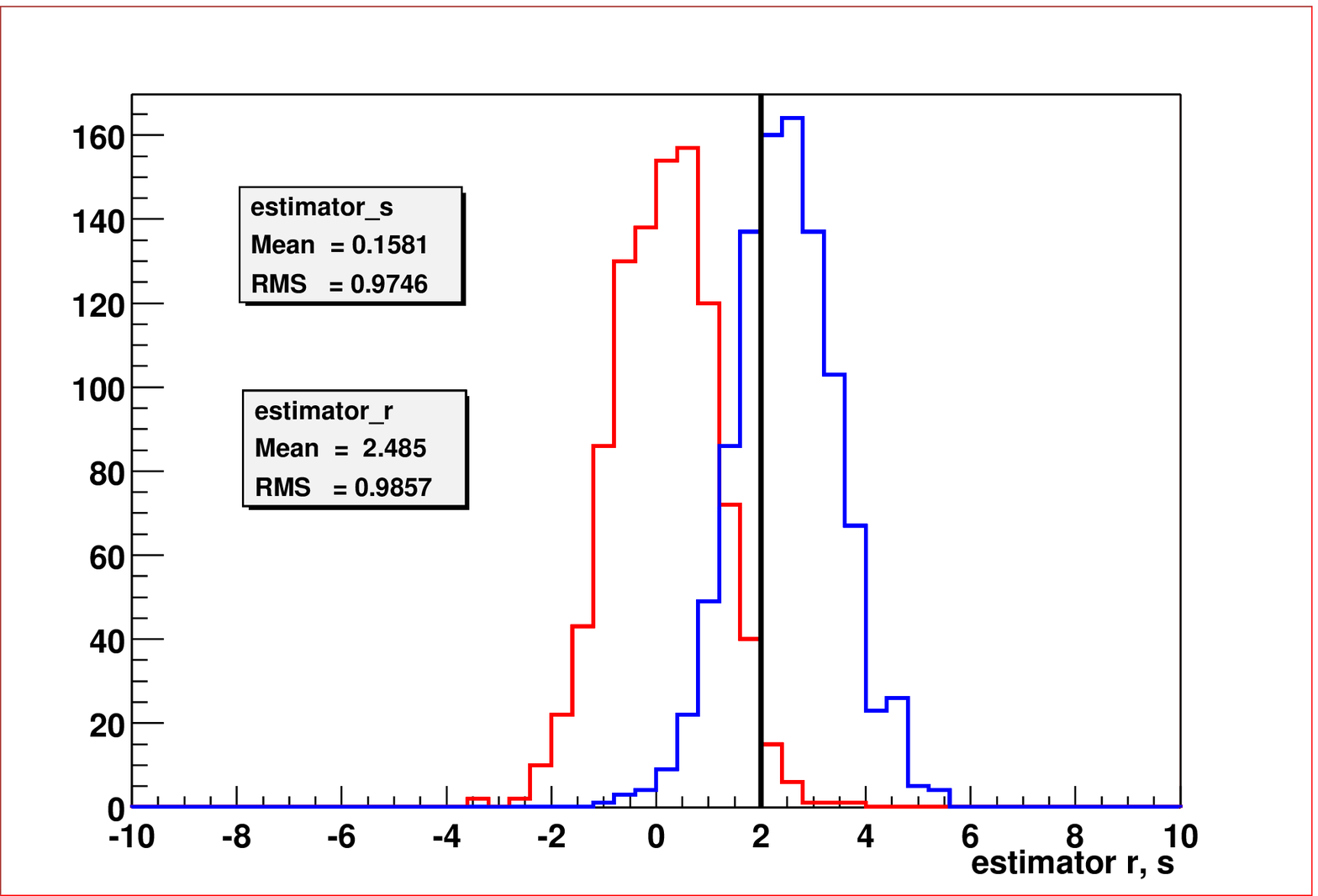}}
%%%\begin{center}
%%%\includegraphics[width=8cm,height=6cm]{est_rs.eps}
%%%\end{center}
\caption{Distributions of the estimators $r$ (blue histogram) and $s$
  (red histogram) for $10^3$
  simulated experiments in the case: $m_W$ = 40 GeV and 
  $\sigma_{\rm{p}} = 5.4\cdot 10^{-6}$ pb. The measuring time is 2
  years and the energy interval where the signal is integrated is
  from 4 keV to 50 keV.}\label{estimator}
\end{figure}

	In fig. 
(\ref{estimator}) 
	we report the distributions of the values for
	the estimators $r$ and $s$ that would be measured in $10^3$ experiments
	like GENIUS-TF after two years of measurement, in the case of a WIMP 
	with mass of 40 GeV and $\sigma_{\rm{p}} = 5.4\cdot 10^{-6}$ pb.

\begin{figure}[!htp] 
\epsfysize=70mm\centerline{\epsffile{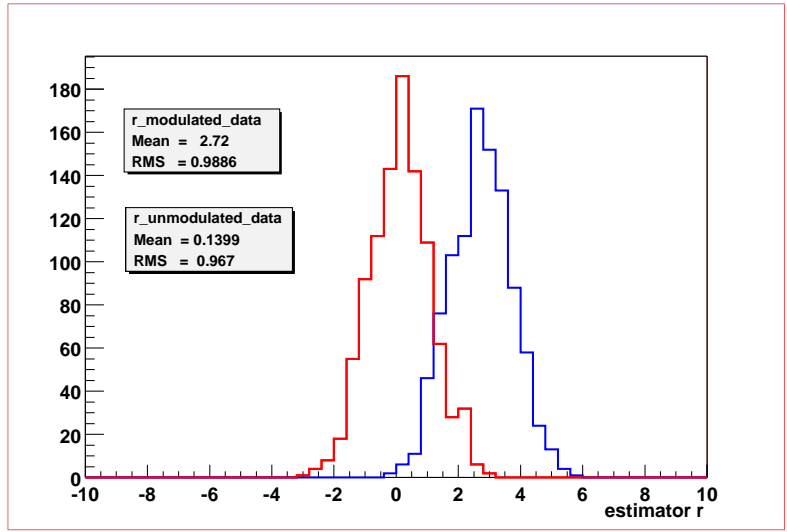}}
%%%\begin{center}
%%%\includegraphics[width=8cm,height=6cm]{est_rs_2.eps}
%%%\end{center}
\caption{Distributions of the estimator $r$ for $10^3$
  simulated experiments in the case of modulated (blue histogram, $m_W$ = 40 GeV and 
  $\sigma_{\rm{p}} = 5.4\cdot 10^{-6}$ pb) and unmodulated data (red histogram). 
  The measuring time is 2
  years and the energy interval where the signal is integrated is
  (4-50) keV.} \label{estimator2}
\end{figure}

	As it is easy to see, the distribution of the variable $s$ is the
	expected Gaussian centered around zero and with unit variance, 
	while the $r$ distribution is shifted to higher values 
	of the variable, indicating the the presence of the modulation. 
	The distribution of the variable $r$ under the same assumptions 
	but without assuming the modulation (only background) is shown in
	fig. 
(\ref{estimator2}), 
	again compared to the distibution under the
	hypothesis of modulation and, as expected, it shows no sign of a
	deviation from the zero mean. 

	For our purposes is interesting to consider the distribution for the 
	modulated data-set. Measuring a value of $r$ equal to the
	mean value of the distribution in fig. 
(\ref{estimator}) 
	will allow us to reject the hypothesis of no modulation 
	with a confidence level $C.L.=0.99\%$.

\begin{figure}[!hbp] 
\epsfysize=70mm\centerline{\epsffile{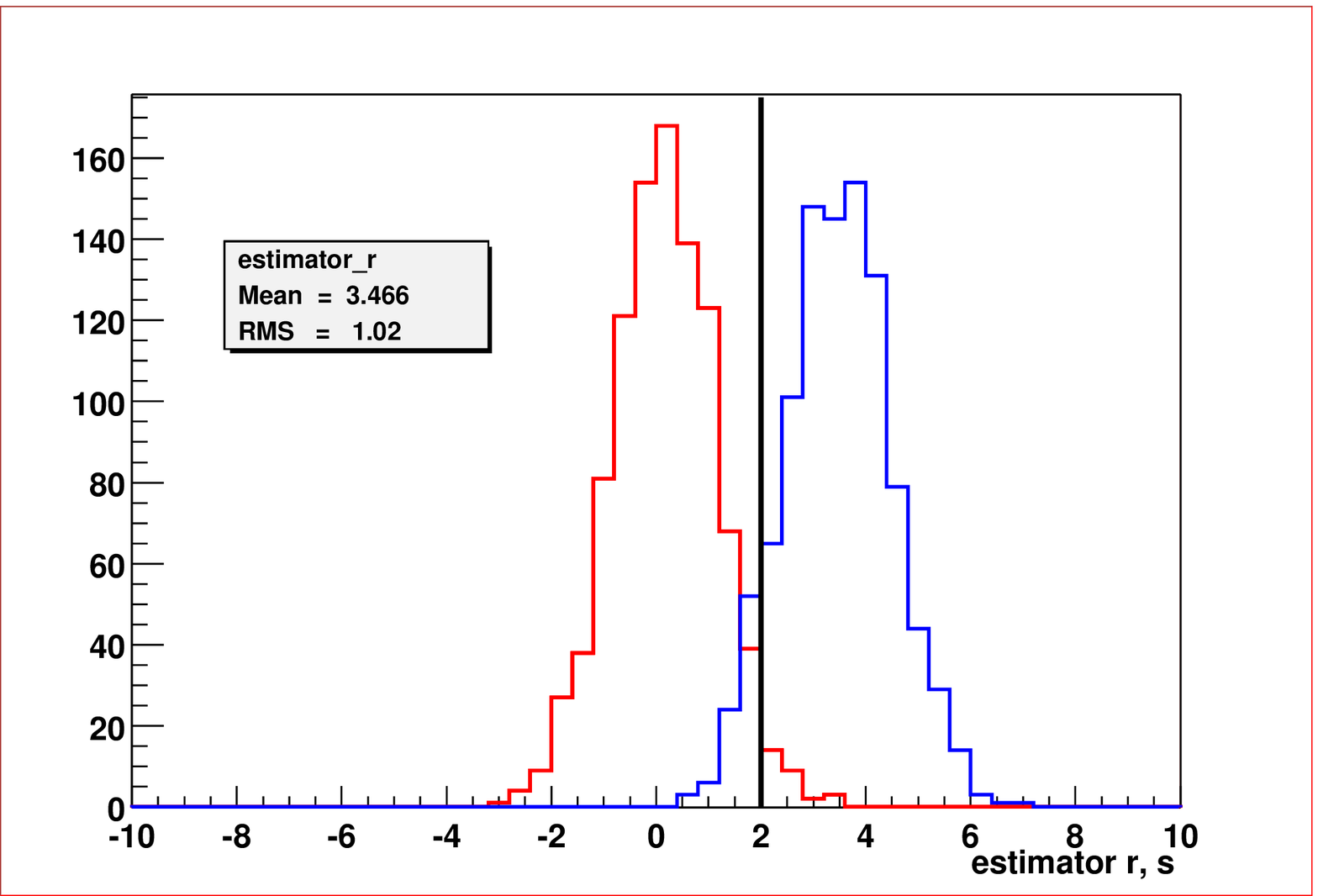}}
%%%\begin{center}
%%%\includegraphics[width=8cm,height=6cm]{est_rs_3.eps}
%%%\end{center}
\caption{Distributions of the estimators $r$ and $s$ for $10^3$
  simulated experiments in the case: $m_W$ = 40 GeV and 
  $\sigma_{\rm{p}} = 5.4\cdot 10^{-6}$ pb. The measuring time is 4
  years and the energy interval where the signal is integrated is
  from 4 keV to 50 keV.} \label{estimator3}
\end{figure}

%%%\noindent
	Obviously, once we set a value for the confidence 
	level $\alpha$ at wich we
	reject the hypotesis of no modulation, there is still 
	a probability to fail
	in detecting the modulation at a confidence level $(1-\beta)$, 
	depending on the overlap region between the two distributions 
	(see 
\cite{Ramachers}). 
	If for example, referring to fig. 
(\ref{estimator}),
	we set $\alpha = 97.5$ ($r_0 = 2$) we will report 
	a positive evidence of 
	the modulation only if the measured value of $r_0$ is greater than 2
	and this correspond to an error of the second kind $(1-\beta)$ = 35\%
	(integral of the $r$ distribution on the interval (0,2)).
	If we want to have a lower error (for the same value of
	$\alpha$) we should wait until the two distributions are separated
	enough, as for example in fig. 
(\ref{estimator3}), 
	with the same assumptions as in 
(\ref{estimator}) 
	and considering 4 years of measurements. 
	In this case we will fail in detecting the modulation
	hidden in the data only in a small fraction of the experiments 
	($1-\beta$ = 8\%).

\begin{table}[!htp] 
\begin{center}
\begin{tabular}{|c|c|c|c|}
\hline
%& & &\\
$m_W$ &  years & $\langle r \rangle$ & C.L. \\
%& & & \\
\hline
50 GeV & 2 & 2.60 & 99.2 \%\\
50 GeV & 4 & 3.74 & 99.9 \%\\
60 GeV & 2 & 2.36 & 98.8 \%\\
60 GeV & 4 & 3.37 & 99.9 \%\\
%& & &\\  
\hline
\end{tabular}
\vspace{0.3cm}
\caption{Mean value of the modulation significance $r$ obtained in
  $10^3$ simulated experiments for different WIMP masses and measuring 
  time in years. The last column contains the confidence level at
  which we could exclude the absence of the modulation when a value of 
  $r$ equal to $\langle r \rangle$ is measured.}
\label{tab2}
\end{center}
\end{table}

%%%\noindent
	The distribution of $r$ for the modulated data-set shows no substantial
	va\-riation when we move the left limit of the energy interval where we
	integrate the signal ($E_i$) from 4 keV to 8 keV. Increasing $E_i$
	above 8 keV, the mean value of the distribution slowly decreases; 
	this is an indication that we are missing part of the signal that,
	as we recall here, is concentrated in the lower energy bins.   
	For this reason is of primary importance to have a low experimental 
	threshold.

	We have repeated the same simulation for different WIMP masses and the 
	same cross section; in tab. 
(\ref{tab2}) 
	we report, for different measuring times, the mean value of the $r$
	distribution (for the modulated data-set) together with the
	corresponding confidence level $\alpha$.

%%%%%%%%%%%%%%%%%% Section 4.2 %%%%%%%%%%%%%%%%%%%%%%%%
\subsection{Extracting the modulated amplitude $S_{\rm{m}}$}
%%%%%%%%%%%%%%%%%% Section 4.2 %%%%%%%%%%%%%%%%%%%%%%%%

	Calculating the value of the above mentioned estimators $r$ and $s$ is 
	not the only way to discover the modulation hidden in the data.
	Given our set of experimental data, we can try to extract directly 
	the value of $S_{\rm{m}}$, the amplitude of the modulated signal, as
	a function of the energy. 

	Following 
\cite{Freese}, 
	we can write:
\begin{eqnarray} \label{Sm}
S_{\rm{m}} &=& \frac{\sum_i 2 \cos\omega (t_i-t_0)S_i}{N} \\
\sigma(S_{\rm{m}}) &=& \frac{\sqrt{\sum_i 2 [\cos\omega (t_i-t_0)]^2 S_i}}{N}  
\end{eqnarray}
	where N is the running time of the experiment and $S_j$ is no
	more integrated over a broad energy region but in 
	small energy bins (1 or 2 keV). In this way it is possible to measure
	the amplitude of the modulated signal and its error for each energy
	bin (of the given size) and moreover, taking small energy bins
	we avoid the problem of the cross-over energy.

\begin{figure}
\epsfysize=70mm\centerline{\epsffile{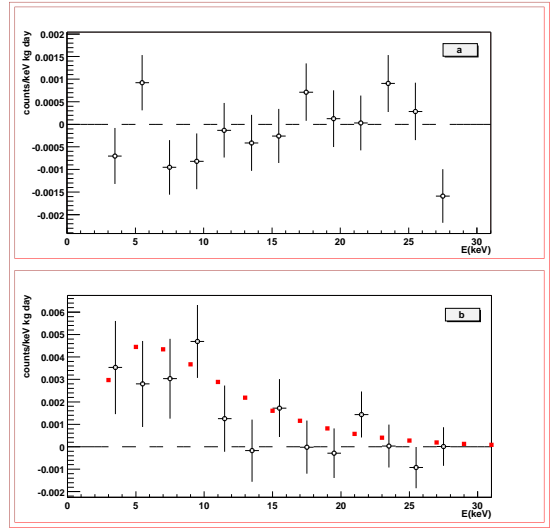}}
%%%\begin{center}
%%%\includegraphics[width=8cm,height=6cm]{Sm_picture.eps}
%%%\end{center}
\caption{Amplitude of the modulated signal extracted from the
  simulated data using eqn.(\ref{Sm}): a) when no modulation is
  present in the data; b) when a WIMP is assumed with $m_W$ = 40 GeV and 
  $\sigma_{\rm{p}} = 5.4\cdot 10^{-6}$ pb. The full squares in
  picture (b) represent the theoretical signal for the given WIMP mass and
  cross section. Both the pictures correspond to a measuring time of 2 
  years.}\label{Sm_picture}
\end{figure}

	In picture 
(\ref{Sm_picture}) 
	we report a tipical result for the amplitude of the modulated 
	signal $S_{\rm{m}}$ as calculated using 
(\ref{Sm}) 
	on the simulated data without and with modulation 
	($m_W$ = 40 GeV and $\sigma_{\rm{p}} = 5.4\cdot 10^{-6}$ pb). 
	The energy region between 4 and 50 keV has
	been divided in 2 keV energy bins and for each 
	one the value of $S_{\rm{m}}$
	and the corresponding error have been calculated. 

	By looking at the two pictures (note the different scales
	on the $y$ axis), we see how the distribution b represents an
	indication of an annual modulation amplitude $S_{\rm{m}} \ne 0 $.
	In the case with
	modulation (b) all the data points are 
	compatible within the error with the theoretical signal while in the
	case where no modulation is present (a) the same points are distributed
	around zero. The big errors are due to the low statistic 
	and become smaller if we increase the measuring time. 

	Averaging the signal in the region from 4 to 16 keV, we obtain
	in the case of modulation: $\langle S_{\rm{m}}\rangle  = 0.0024
	\pm 0.0016$ counts/keV day, while in the case of no modulation we
	have: $\langle S_{\rm{m}}\rangle  =  -0.00034 \pm 0.0006$ counts/keV
	day, compatible with the hypothesis $S_{\rm{m}} = 0 $.

%%%%%%%%%%%%%%%%%% Section 4.3 %%%%%%%%%%%%%%%%%%%%%%%%
\subsection{The maximun likelihood method}
%%%%%%%%%%%%%%%%%% Section 4.3 %%%%%%%%%%%%%%%%%%%%%%%%

	One of the general procedures to treat experimental 
	data when searching 
	for the annual modulation effect is to use the maximum-likelihood
	method. The likelihood function $L$ for a set of experimental rates
	$N_{ij}$, assuming that they are Poisson-distributed, is:
\begin{equation} \label{Lik}
L = \prod_{ij} e^{-\mu_{ij}}\frac{\mu_{ij}^{N_{ij}}}{N_{ij}!}.
\end{equation}
	Since $\mu_{ij}$ depends on the parameters of interest $m_W$ and
	$\sigma_p$, through the expressions of $S_{0,j}$ and $S_{m,j}$, it is
	possible to obtain the best-fit value for the parameters minimizing
	$\log L$, or better the function:
$$
y(m_W,\sigma_p)=-2\log L + {\rm{const}} = \sum_{ij} 2\mu_{ij} -
2N_{ik}\log\mu_{ij} + {\rm{const}}
$$
	where the constant contains the components that do not depend on 
$m_W$ and $\sigma_p$.
	The minimization of $y(m_W,\sigma_p)$ is not an easy operation: if we
	want to keep as a free parameter the number of background counts in
	each energy bin (and here we consider as background 
	the number of counts due to the time-independent component 
	of the signal ($S_0$ + $b$))
	we have to deal with many parameters. 

	The usual procedure is to carry out
	the minimization in two steps. 
	As a first step we minimize with respect to the
	time-independent component $f_j = b_j + S_{0,j}$. During this
	minimization the condition $f_j >0$, $j=1,N_{bin}$ is imposed.
	As a second step we minimize with respect to $\sigma_p$ 
	and $m_W$ requiring, as is usual done:
\begin{eqnarray*}
(b_j + S_{0,j}) &=& f_j \qquad \qquad \mbox{if} \;\; \sigma_p \; S_{0,j} \leq f_j \\
(b_j + S_{0,j}) &=& \sigma_p S_{0,j} \qquad \mbox{otherwise.} 
\end{eqnarray*}
	We did not impose limits or conditions on the parameters $\sigma_p$ 
	and $m_W$ except that both have to be greater than zero.

\begin{figure}
\epsfysize=70mm\centerline{\epsffile{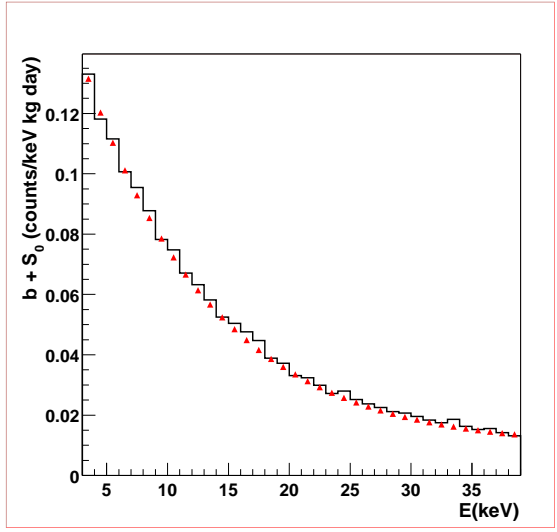}}
%%%\begin{center}
%%%\includegraphics[width=8cm,height=6cm]{f_k.eps}
%%%\end{center}
\caption{The time-independent component of the signal $f = b + S_0$:
  what we have simulated (red triangles) and what we obtain in
  the first step of the minimization procedure (solid histogram).}\label{f_k}
\end{figure}

	We have carried out the minimization procedure on the data
	set simulated under the mentioned hypothesis $m_W$ = 40 GeV and 
	$\sigma_{\rm{p}} = 5.4\cdot 10^{-6}$ pb, in the energy range from 5 to 
	40 GeV. After the first step we
	obtain the values of the quantity $f_j = b_j + S_{0,j}$ for each
	energy bin: we have plotted that result in fig. 
(\ref{f_k}) 
	and this reproduces exactly the curve $f = b + S_0$ that we have 
	used in the simulation.

	In the second step we obtain the best-fit values for 
	the parameters of interest.

	The minimization procedure converged for the selected 
	energy window, gi\-ving the following best-fit values for the fit
	parameters:
\begin{eqnarray} \label{best-fit}
m_W &=& (39.9 \pm 5.6) \; \rm{GeV} \nonumber \\
\sigma_p &=& (7.0 \pm 1.6)\cdot 10^{-6} \; \rm{pb} 
\end{eqnarray}
	The value of the WIMP mass is in excellent agreement with the real one 
	used in the sumulation; the value of the cross section, though
	compatible within 1 $\sigma$ with the true value, shows a lower
	agreement: this can be due to the fact that $m_W$ influences the
	shape of the WIMP spectrum while $\sigma_p$ appears only as a
	multiplying factor.
 
	We repeated the fitting procedure assuming different values 
	of the energy threshold of the experiment (3 and 4 keV), 
	finding very little difference from the given best-fit. 

\begin{figure}
\epsfysize=70mm\centerline{\epsffile{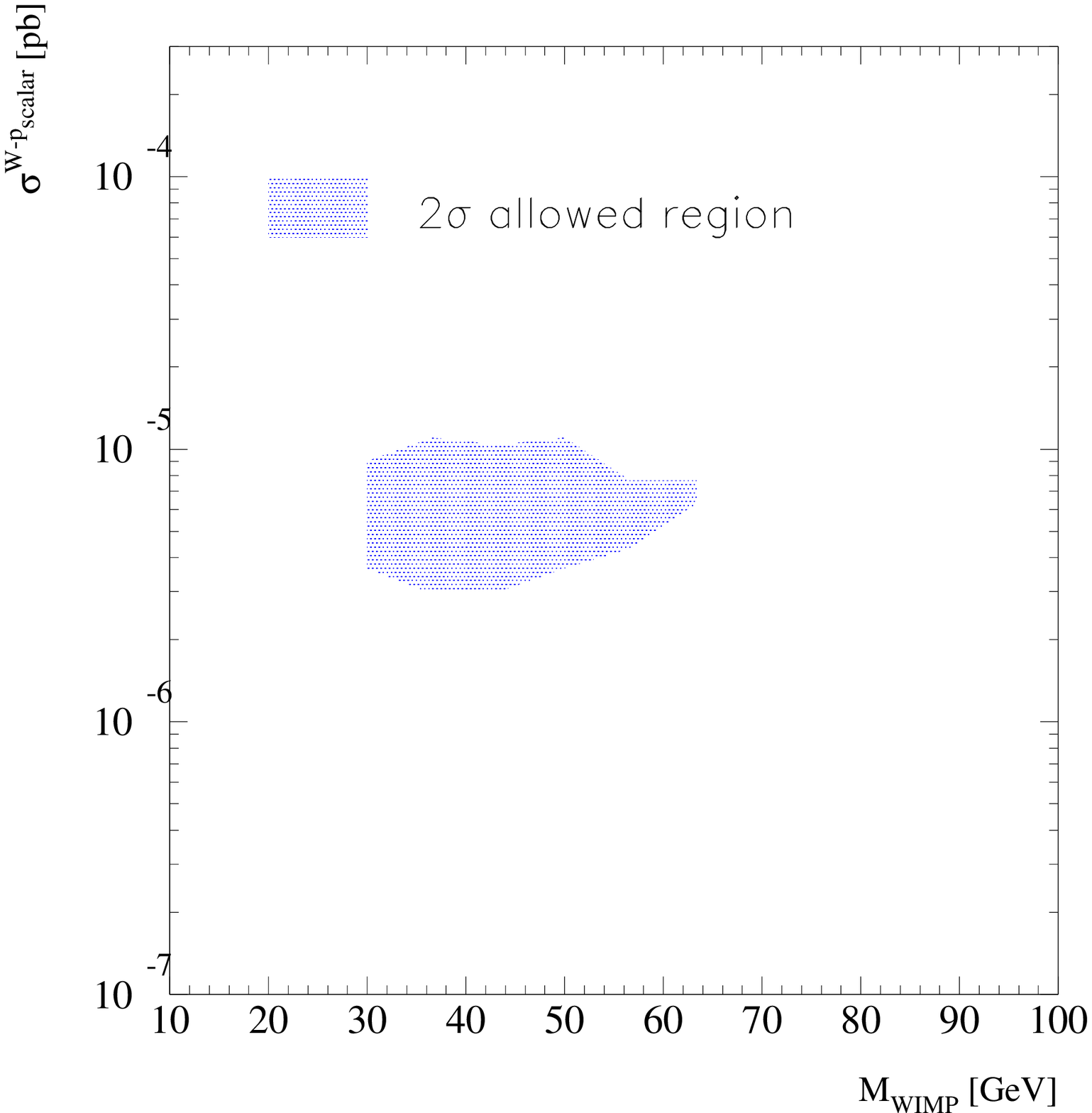}}
%%%\begin{center}
%%%\includegraphics[width=8cm,height=7cm]{region.eps}
%%%\end{center}
\caption{Allowed region at $2\sigma$ C.L. corresponding to the
  best-fit values of (\ref{best-fit}). The region is calculated 
  using eq.(\ref{excl}) and has to be interpreted as the result that can 
  be given by GENIUS-TF after two years of measurement 
  if a WIMP exists with the properties assumed 
  so far.} \label{region}
\end{figure}

	The set of best-fit values (\ref{best-fit}) corresponds 
	to the 2 $\sigma$ allowed region shown in fig. 
(\ref{region}). 
	This region of the plane $(\sigma_p,m_W)$ has been calculated using: 
\begin{equation} \label{excl}
y(\sigma_p,m_W)-y_{\rm{min}} \le n^2 .
\end{equation}
	with $n=2$ and $y_{\rm{min}}$ being the result 
	of the fitting procedure.\\
	Fig. 
\ref{region} 
	should be interpreted in this way:
	if a WIMP exists with the properties assumed so far, we 
	can point out its presence within two years of measurement (80 kg y
	significance) and give the result in Fig. 
\ref{region} 
	for the 2 $\sigma$ allowed region for the relevant parameters.

%%%%%%%%%%%%%%%%%% Section 5 %%%%%%%%%%%%%%%%%%%%%%%%
\section*{Conclusions}
%%%%%%%%%%%%%%%%%% Section 5 %%%%%%%%%%%%%%%%%%%%%%%%

	The annual modulation, due to the motion of the earth with respect to
	the galactic halo, is the main signature of a possible WIMP signal.
	The effect is supposed to be small, only a few percent of the total
	Dark Matter signal, and therefore very difficult to extract.
	A positive indication of this modulation has been found 
	over the past five years by the DAMA experiment 
	and it would be of great importance to
	look for the same effect with another experiment, expecially in the
	region of the WIMP parameter's space indicated by the DAMA results. 

	The GENIUS-TF experiment 
\cite{GENIUS-TF}, 
	born as a prototype for the GENIUS project,
	is at present under installation at the Gran Sasso National
	Laboratory. With a mass of 40 kg and a background of 4 counts/(kg y
	keV), GENIUS-TF can be used to look for Dark Matter, 
	not only through the
	direct detection of WIMP-induced nuclear recoils, but also through the
	annual modulation of the experimental rate. 
	GENIUS-TF will be - in addition to DAMA
\cite{DAMA03}
	 - the {\it only} 
	experiment which will be able to probe the annual modulation 
	signature in a foreseeable future. 
	The at present much discussed cryo detector experiments, 
	such as CDMS
\cite{CDMS02}, 
	CRESST 
\cite{CRESST02}, 
	EDELWEISS 
\cite{EDELWEISS02}
	have no chance to do this because the mass projected 
	to be in operation in these experiments is by far too low (see also 
\cite{KK-LP01}).

	We have developed a set of routines and tools to look for the annual
	modulation effect with our GENIUS-TF experiment using different
	analysis methods discussed in the literature.    
	We have analysed data simulated under the hypothesis of modulation, 
	using three different approaches (modulation significance, direct
	calculation of the modulation amplitude $S_m$ and maximum likelihood
	fit of the data)  and we have shown that the
	mass and the low background level of GENIUS-TF will allow us test,
	within few years of measurements, low WIMP masses and 
	WIMP cross sections in the region of interest indicated 
	by the DAMA experiment.
	A digital multi-channel spectroscopy system with 100\,MHz 
	flash ADC's for the GENIUS-TF project has been recently developed 
\cite{KK-Electr-TF03}.

%%%%%%%%%%%%%%%%%%%%%%% section 6 %%%%%%%%%%%%%%%%%%%%%%%

%%%%%%%%%%%%%%%%%%%%%%% end section 5 %%%%%%%%%%%%%%%%%%%%%%%

%%%%%%%%%%%%%%%%%%%%%%% FIGURES %%%%%%%%%%%%%%%%%%%

\clearpage
%%%%%%%%%%%%%%%%%% Fig. 1 %%%%%%%%%%%%%%%%%%%%
%%%%%%%%%%%%%%%%%% Fig. 1 %%%%%%%%%%%%%%%%%%%%

%%%%%%%%%%%%%%%%%% Fig. 2 %%%%%%%%%%%%%%%%%%%%
%%%%%%%%%%%%%%%%%% Fig. 2 %%%%%%%%%%%%%%%%%%%%

%%%%%%%%%%%%%%%%%% Fig. 3 %%%%%%%%%%%%%%%%%%%%
%%%%%%%%%%%%%%%%%% Fig. 3 %%%%%%%%%%%%%%%%%%%%

%%%%%%%%%%%%%%%%%% Fig. 4 %%%%%%%%%%%%%%%%%%%%
%%%%%%%%%%%%%%%%%% Fig. 4 %%%%%%%%%%%%%%%%%%%%

%%%%%%%%%%%%%%%%%% Fig. 5 %%%%%%%%%%%%%%%%%%%%

%%%%%%%%%%%%%%%%%% Fig. 5 %%%%%%%%%%%%%%%%%%%%

%%%%%%%%%%%%%%%%%% Fig. 6 %%%%%%%%%%%%%%%%%%%%
%%%%%%%%%%%%%%%%%% Fig. 6 %%%%%%%%%%%%%%%%%%%%

%%%%%%%%%%%%%%%%%% Fig. 7 %%%%%%%%%%%%%%%%%%%%
%%%%%%%%%%%%%%%%%% Fig. 7 %%%%%%%%%%%%%%%%%%%%

%%%%%%%%%%%%%%%%%% Fig. 8 %%%%%%%%%%%%%%%%%%%%
%%%%%%%%%%%%%%%%%% Fig. 8 %%%%%%%%%%%%%%%%%%%%

%%%%%%%%%%%%%%%%%% Fig. 9 %%%%%%%%%%%%%%%%%%%%
%%%%%%%%%%%%%%%%%% Fig. 9 %%%%%%%%%%%%%%%%%%%%


\begin{thebibliography}{99}

\bibitem{Freese} 
	A.K. Drukier, K. Freese and D.N. Spergel, 
	Phys. Rev. D 33 (1986) 3495.

\bibitem{Freese2} 
	K. Freese, J. Friedman and A. Gould, 
	Phys. Rev. D 37 (1988) 3388.

\bibitem{DAMA} 
	R. Bernabei et al., Phys. Lett. B 480 (2000) 23.

\bibitem{cross1} 
	R. Bernabei et al., Phys. Lett. B 389 (1996) 757.

\bibitem{cross2} 
	J.D. Lewin, P.F. Smith, Astropart. Phys. 6 (1996) 87.

\bibitem{Baudis} 
	L. Baudis, Diploma Thesis, University of Heidelberg (1997).

\bibitem{Cebrian} 
	S. Cebri\'an et al., Astropart. Phys. 14 (2001) 339.

\bibitem{GENIUS-TF} 
	H.V. Klapdor-Kleingrothaus et al., 
	Nucl. Instrum. Meth. A481 (2002) 149-159. 

\bibitem{GENIUS} 
	H.V. Klapdor-Kleingrothaus, J. Hellmig and M Hirsch,
  J.Phys.G: Nucl. Part. Phys. 24 (1998) 483-516;
  H.V. Klapdor-Kleingrothaus, L. Baudis, G. Heusser, 
  B. Majorovits and H. P\"as, ``GENIUS: a Supersensitive Germanium 
  Detector System for Rare Events'', Proposal MPI-H-V26-1999,
  hep-ph/9910205 and in Proceedings of the Second International
  Conference on Particle Physics beyond the Standard Model, BEYOND THE
  DESERT 1999, Castle Ringberg, Germany 6-12 June 1999, ed. by 
  H.V. Klapdor-Kleingrothaus, I. Krivosheina (IOP Bristol 2000), 915.

\bibitem{Bela1} 
	B. Majorovits, PhD Thesis, University of Heidelberg (2000).

\bibitem{Bela2} 
	B. Majorovits et al., 
	Nucl. Instr. Meth. A 455 (2000) 369.

\bibitem{DAMA2} 
	R. Bernabei et al., Phys. Lett. B 424 (1998) 195.

\bibitem{Hasenbalg} 
	F. Hasenbalg, Astropart. Phys. 9 (1998) 339.

\bibitem{Sarsa} 
	M.L. Sarsa et al., Phys. Rev. D 56 (1997) 1856.

\bibitem{Ramachers} 
	Y. Ramachers, M. Hirsch and H.V. Klapdor-Kleingrothaus, 
	Eur. Phys. J. A3 (1998) 93.

\bibitem{KK-Electr-TF03}
	T. Kihm, V.F. Bobrakov and H.V. Klapdor-Kleingrothaus,
	to be publ. in NIM (2003).

\bibitem{KK-LP01}
	H.V. Klapdor-Kleingrothaus, 
	{\it Int. J. Mod. Phys.} {\bf A17}, 3421-3431 (2002), 
	in Proc. of {\it Internat. Conf. LP01}, Rome, Italy, July 2001. 

\bibitem{DAMA03}
	R. Bernabei et al., in {\it Proc. of Intern. Conf. 
	on Physics Beyond the Standard Model: 
	Beyond the Desert 02, BEYOND'02}, Oulu, Finland, 2-7 Jun 2002, 
	{\it IOP, Bristol}, 2003, ed. H.V. Klapdor-Kleingrothaus.

\bibitem{CDMS02}
	T. Saab et al., 
	Nucl. Phys. Proc. Suppl. 110 (2002) 100-102.

\bibitem{CRESST02}
	F. Probst et al., 
	Nucl. Phys. Proc. Suppl. 110 (2002) 67-69.

\bibitem{EDELWEISS02}
	A. Benoit et al., 
	Phys. Lett. B 545 (2002) 43-49.


\end{thebibliography}
\end{document}